\documentclass[aps,showpacs,dvips,showkeys,preprint,preprintnumbers]{revtex4}

\usepackage{amsmath}
\usepackage{amssymb}
\usepackage{graphicx}
\usepackage{color}

\def \be  {\begin{equation}}
\def \ee  {\end{equation}}
\def \ee  {\end{equation}}
\def \bea {\begin{eqnarray}}
\def \eea {\end{eqnarray}}

\newcommand{\nn}{\nonumber}

\begin{document}

\preprint{ECTP-2016-16}
\preprint{WLCAPP-2016-16}
\hspace{0.05cm}
\title{Minimal-supersymmetric extended inflation field in Horava-Lifshitz gravity}

\author{Abdel Nasser Tawfik}
\email{a.tawfik@eng.mti.edu.eg}
\affiliation{Egyptian Center for Theoretical Physics (ECTP),  MTI University, 11571 Cairo, Egypt}
\affiliation{World Laboratory for Cosmology And Particle Physics (WLCAPP), 11571 Cairo, Egypt}

\author{Abdel Magied Diab}
\affiliation{Egyptian Center for Theoretical Physics (ECTP),  MTI University, 11571 Cairo, Egypt}
\affiliation{World Laboratory for Cosmology And Particle Physics (WLCAPP), 11571 Cairo, Egypt}

\author{Eiman Abou El Dahab}
\affiliation{Faculty of Computer Sciences, MTI University, 11671 Cairo, Egypt}
\affiliation{World Laboratory for Cosmology And Particle Physics (WLCAPP), 11571 Cairo, Egypt}

\date{\today}

\begin{abstract}
We study the Friedmann inflation in general covariant Horava-Lifshitz (HL) gravity without the projectability conditions and with detailed and non-detailed balance conditions. Accordingly, we derive modifications in the Friedmann equations due to a single-scalar field potential describing minimal-supersymmetrically extended inflation. By implementing two time-independent equations of state (EoS) characterizing the cosmic background geometry filled up with dark energy, the dependence of the tensorial and scalar density fluctuations and their ratios on the inflation field are determined. The latter refer to the time evolution of the inflationary field relative to the Hubble parameter. Furthermore, the ratios of tensorial-to-spectral density fluctuations are calculated in dependence on the spectral index. For cold dark energy EoS $\omega=-1/3$, we find that the tensorial-to-spectral density fluctuations are not depending on the different theories of gravity and the results are very small relative to the recent BICEP2/Keck Array-Planck observations, $10^{-9} \lessapprox r \lessapprox 10^{-3}$. We have also calculated the tensorial and scalar perturbations of the primordial spectra. 
\end{abstract}


\maketitle


\section{Introduction\label{sec:intr}}

There are various theories suggesting proposals to solve specific problems arising from the general relativity gravity (GRG) \cite{nori,masud}. It was shown that,  they likely unify different inflation models with the cosmological observations on the late-time acceleration. In the strong gravitational regimes, Horava-Lifshitz (HL) gravity was assumed as an estimate for the possible modifications to GRG \cite{lifshitz,Horava2009a}. HLG explains well the expansion of our-apparently-accelerated Universe \cite{hlg2010}. Thus, we propose to implement HLG, especially to the inflation era.

For the scale-invariant quantum fluctuations,  a plausible scenario based on HLG has been proposed \cite{0904.2190}. Concretely, this implements HLG without the additional scalar degree-of-freedom (dof) and accordingly an inflation era is produced in a very early stage of the Universe. Although, no need for the detailed balance conditions because of the horizon problems, but we require that the inflation itself should be possessing slow-roll parameters. In the present work, we plan to introduce a single scalar field potential describing the minimal-supersymmetrically extended inflation and leading to modifications in the Friedmann equations.

In HLG without the projectability conditions \cite{Wang2011sds, Wang2012ss} and with the projectability conditions \cite{daSilva2011ds,Malik2009ad}, the spin-$0$ gravitons could be eliminated though implementing a local U($1$) symmetry \cite{Zhu2012}, where various problems concerning instability, strong coupling and different speeds in the gravitational sector are conjectured to be solved \cite{Wang2011sds}.  Relative to HLG without the projectability conditions, imposing the detailed balance conditions significantly reduces the number of independent coupling constants \cite{Wang2011sds}. Furthermore, when implementing HLG with and without the projectability conditions in a local U($1$) symmetry to the cosmology, it was found that the Friedmann-Lemaitre-Robertson-Walker (FLRW) Universe becomes flat \cite{Wang2012ss,Huang2011aa}. The linear perturbations of the scalar and tensor perturbations of the resulting Universe were investigated, as well. For instance, for the scalar perturbations, it was found that the field equations are almost similar to those for the Einstein's theory of general relativity  \cite{Wang2012ss,Malik2009ad}. For the tensor perturbations, no contributions are came up with for the gravitational and matter sectors \cite{Wang2012ss}. Also, we noticed that when applying HLG to the FLRW Universe various problems are avoided, such as, (i) at low energy, although GRG is not fully recovered but imitates it with an additional dark matter sector,  (ii) its higher spatial curvature terms seem to allow for scenarios such as bouncing and cyclic Universe, i.e. avoiding the sever Big Bang singularity, and (iii) through its anisotropic scaling the horizon problem and even more the scale-invariant cosmological perturbations are generated even without an inflationary epoch \cite{Calcagni2009ss,Brandenberger2009ss,Mukohyama2010}. The static diagonal vacuum solutions of the HLG without the projectability conditions leads to very rich Lifshitz-type structures including the Lifshitz spacetimes with or without hyperscaling violation, Lifshitz solitons, and black holes, seems to allow generically instantaneous propagations, e.g. universal horizons still exist, which serve as one-way membranes for signals with any large velocities, and explicitly breaks the Lorentz symmetry \cite{Wen2015ssa}.

In HLG without the projectability conditions, the inflationary Universe has been extensively studied \cite{Wang2012ss,1007.1006,1004.2552}. For a single scalar field, linear scalar perturbations to the FLRW equations have been derived \cite{1007.1006,1004.2552}. For the sake of completeness, we emphasize that this result is just the opposite to the one reported about in Ref. \cite{0904.2190}. For a gauge spectrum index and power spectrum of comoving curvature, the quantum perturbations have been estimated \cite{Baumann2009}. The HLG without an additional scalar degree-of-freedom, i.e. no inflationary epoch, was proposed as a simple scenario for the scale-invariant quantum-fluctuations \cite{Mukohyama2010}. For this proposal, it was found that the detailed balance conditions are not necessary, but serious horizon problems, such as monopole and domain walls remain unsolved. Solving them requires inflation with slow-roll conditions. The present work suggests an opposite scenario, i.e. an inflationary epoch. Contrary to the proposal of Ref. \cite{Mukohyama2010}, the linear scalar perturbations equations of the FLRW Universe can be derived for a single scalar field. At various {\it time-independent} equations of state (EoS), the power spectrum and spectrum index of the comoving curvature perturbations shall be determined \cite{Tawfik:2016dvd}. Even in HLG, including scalar fields is essential for the inflation. It was found that the perturbations are scale-invariance and  HLG becomes consistent with various cosmological observations. This phenomenological study gives an additional support for imposing scalar field(s), e.g. inflation. Recently, a most recent introduction and progresses in HLG were outlined and reviewed \cite{Wang2017}. 

We consider two sorts of cold dark energy, an exotic energy form with negative pressure and thus likely responsible for the cosmic acceleration, namely {\it time-independent} EoS ($\omega=p/\rho $), where $\omega=-1/3$ and $-2/3$. For the sake of simplicity, we assumed that the EoS are time-independent. The reason is obvious. This shall allow us highlighting the importance of our proposal of the inclusion of scalar fied(s), e.g. inflation, which represents an alternative scenario to the one presented in Ref. \cite{Mukohyama2010}. In a future work, we shall work out calculations at time-dependent EoS \cite{NFS2012}. In this case, the system of interest is likely out of equilibrium, which makes it more complicated to deal with. The dynamical nature of the dark energy could be understood from various fields characterized through quantum gravity. The earlier EoS refers to non-relativistic cosmic strings and curvature acting as a density component, where $\rho \propto a^{-2}$. It is noteworthy remarking that networks of cosmic strings possess a similar scaling. The curvature driven Universe, known as Milne Universe, linearly expands without acceleration or deceleration. The latter EoS, i.e. $\omega=-2/3$, describes non-relativistic domain walls and thus refers to cosmological acceleration, where the scalar fields are known as quintessence are phantom fields. So far, there is no observation for the dynamical evolution of the dark energy, except for  the cosmological constant ($\omega=-1$).  

For the sake for completeness, we recall that, the projectability condition is known to cause many problems such as unitary breaking, strong coupling, and gradient instability. Moreover, with projectability, the Friedmann equation derived from the Hamiltonian constraint is no longer an equation of motion, but only its integral over all space is. 
 
The present letter aims to combine recent cosmological observations with the possible distinguishability between GRG and HLG. We apply a cosmic inflation model to HLG and compare between different types of HLG, which are proposed to reduce the coupling constants of HLG. A short reminder to the theory of HLG shall be given in section \ref{sec:HLG}. Also, the possible modifications to the Friedmann equations due to HLG with non-detailed and detailed balance conditions, and without the projectability conditions shall be estimated in sections \ref{sec:HLGndb}, \ref{sec:HLGdb}, and \ref{sec:wtproj}. The results shall be discussed in section \ref{sec:Fried}. The resulting fluctuations and slow-roll parameters followed by a study for the tensorial and scalar perturbations of the primordial spectra shall be elaborated. Our results confronted with the recent PLANCK observations shall be discussed. Last, we outline the final conclusions in section \ref{sec:conc}.

\section{Reminder to Horava-Lifshitz Gravity}
\label{sec:HLG}

For quantum gravity with an anisotropic ultraviolet (UV) scale, Horava scalar field has been proposed \cite{Horava2009a}, where the fundamental assumptions of this theory are related to the Minkowskian space-time. This was inspired by the Lifshitz scalar field for condensed matter theory \cite{lifshitz}. HLG is known as a projectable approach minimizing the number of independent potential couplings and adopting an extra principle in constructing the potential.  The general relativity (GR) is known as emerging in an infrared (IR) fixed point so that in some limits of the dynamical critical exponent ($z$), GR can be recovered \cite{ir1}. HLG exhibits a strong anisotropy between space and time, 
\begin{eqnarray}
t \longrightarrow  l^{z}\, t, \qquad & & x^{i} \longrightarrow  l\, x^{i},
\end{eqnarray}
where $l$ is a constant. As the anisotropic scaling implies a preferred time coordinate, it is obvious that $4$D Lorentz becomes non-invariant (pseudo-Riemannian) manifold. In a differentiable manifold, codimension foliation as its basic structure can be resorted and both spatial and temporal coordinates can be separated  \cite{Horava2009a}.

HLG with an additional scalar field is assumed to generate an effective dark energy \cite{Saridakis:2009bv,Setare:2009vm}. Among cosmologists, it isn't unfamiliar that, in inflationary Universe an {\it ad hoc} EoS for dark energy can be ruled out \cite{1410.6514,1508.01787}. But for late time cosmology where HLG is conjectured to reduce to Einstein's GRG, evidences for dark energy appear. Furthermore, it was found that, the HL dark energy stemming from its additional scalar field seems to possess a varying EoS. At small scale factor, this replaces the problematic singularity with a bounce and at large scale factors it triggers turnaround, i.e. leading to cyclic cosmology \cite{Saridakis:2009bv}. The present genuine scientific work introduces various novelties, a) emphasizes on the tensor to scalar ratio, b) presents application to a particular class of scalar field potentials which are motivated by minimal supersymmetry, and c) reexamines the inclusion of dark energy.

\subsection{HLG with non-detailed balance conditions}
\label{sec:HLGndb}

To reduce the number of independently coupling constants \cite{const1a,const1b,const2}, non-detailed balance conditions were proposed \cite{Horava2009a}, especially when general covariance becomes abandoned \cite{Wang2012ss,1007.1006,1004.2552}. This is the first condition assuming that the gravitational potential can be estimated from a super potential defined on three spatial hypersurfaces, while the fourth dimension $t$ is taken constant. 

Possible modifications to the Friedmann equations \cite{Tawfik:2017ngn,ir1,liu2016,Tawfik:2016dvd,BA,Wei2011,Gomez2011} with HLG with non-detailed balance conditions read 
\bea
H^2 &=& \frac{\kappa^2}{6(3 \lambda -1 )} \rho \,-\,\frac{\kappa^4 \mu^2 \Lambda ^2 _{\omega}}{16(3 \lambda -1)^2} + \frac{\kappa ^4 K \mu ^2 \Lambda _{\omega}}{8(3 \lambda -1 )^2 a^2}-\frac{\kappa ^4 K^2 \mu ^2 }{16(3 \lambda -1)^2 a^4}, \label{Freid1} \\
\dot{H} + \frac{3}{2} H^2 &=& -\frac{\kappa ^2}{4(3 \lambda -1 )} p -\frac{\kappa ^4 \mu ^2 \Lambda ^2 _{\omega}}{32(3 \lambda -1 )^2}  + \frac{\kappa ^4 K \mu ^2 \Lambda _{\omega}}{16(3 \lambda -1 )^2 a^2},
\eea
where $\mu$ and $\Lambda _{\omega}$ are free parameters. They have mass dimensions with $[\mu]=1$ and $[\Lambda _{\omega}]=2$. $\lambda$ is a new dimensionless coupling, whose role in HL theory is still under debate. It is the parameter which measures the departure of the kinetic terms from GR equivalents, as $\lambda=1$. The Hubble parameter $H=\dot{a}/a$, $\dot{H}=\ddot{a}/a - H^2$. The extrinsic curvature of spatial slices  $K=g^{i\,j}\, K_{i\,j}$. $K_{i\,j}=(\partial_t\, g_{i\,j}^{(3)} - \nabla_i^{(3)}\, N_j - \nabla_j^{(3)}\, N_i)/(2 N)$. The Einstein coupling is $\kappa=\sqrt{8\pi G}/c^{2}$, where $G$ is the gravitational constant. In natural units, the speed of light can be omitted; $c=1$. Under this scaling, all variables become dynamical, 
\begin{eqnarray}
N \longrightarrow  N, \quad
g_{ij} \longrightarrow g_{ij},  \quad 
N_{i} \longrightarrow  l^{2}\, N_{i}, \quad
N^{i} \longrightarrow  l^{-2}\, N^{i}.
\end{eqnarray}
The Ricci scalar reads
\begin{eqnarray}
R = K_{i j}\, K^{i j} - K^{2} + R^{(3)} 
+ 2 \nabla_{\mu}\left(n^{\mu} \nabla_{\nu} n^{\nu} - n^{\nu} \nabla_{\nu} n^{\mu}\right),
\end{eqnarray}
where $R^{(3)}$ is the spacial scalar curvature, $n^{\nu}$ is a unit vector perpendicular to a hypersurface representing the time, and $\nabla$ is the covariant derivative on the spatial slice. Slicing  the space-time was detailed in Refs. \cite{Tawfik:2017ngn,ir1,liu2016,Tawfik:2016dvd,BA,Wei2011,Gomez2011}. Accordingly, by using the cosmic time, $g_{ij}=a^2 \delta_{ij}$ and under the assumption that $\xi=\kappa^2/(3 \lambda -1)$, Eq. (\ref{Freid1}) becomes 
\bea
\label{Fred2}
H^2 = \frac{\xi}{6} \rho - \frac{\xi ^2 \mu ^2 }{16} \left[ \Lambda ^2 _{\omega} - \frac{2 K \Lambda _{\omega}}{a^2} + \frac{K^2}{a^4} \right],
\eea
where $\rho$ is the energy density.
The continuity equation can be given as 
\bea 
\rho = \rho_o a^{-3 (1+\omega)},
\eea
where $\omega$ is the proportionality constant relating energy density with pressure, known as EoS. In natural units, $\omega\equiv c_s^2$; the squared speed of sound . This leads to scale factor $a \approx \rho^{-1/[3(1+ \omega)]}$. Then, Eq. (\ref{Fred2}) can be rewritten as
\bea
H^2 = \frac{\xi}{6} \rho - \frac{\xi ^2 \mu ^2 }{16} \left[ \Lambda ^2 _{\omega} - 2 K \Lambda _{\omega} \, \rho ^{\frac{2}{3(1+\omega)}} + K^2 \, \rho ^{\frac{4}{3(1+\omega)}} \right].
\eea

Let us start with the classical field theory. As function of the total inflation energy ($\phi$), the inflation scalar field reads
\bea
\frac{1}{2} \left( \dot{\phi}^2 + (\nabla \phi)^2\, \right) + V(\phi). \label{infl_ener}
\eea
This field is coupled to gravity. The dynamics of the inflation can be characterized as follows.
\begin{itemize}
\item In homogeneous and isotropic universe, the Friedmann equation describes the expansion and contraction such as
\bea
H^2 + \frac{k}{a^2}&=& \frac{8\, \pi\, G\, \rho}{3} + \frac{\Lambda}{3},
\eea
where $k=[+1,-1,0]$ stand for open and closed curved and flat Universe. 
\item For a spatially homogeneous scalar field, the Klein-Gordon equation represent an EoS,
\bea
\ddot{\phi} + 3 \,H\, \dot{\phi} + \partial_\phi\, V(\phi) = 0. \label{KG}
\eea
\end{itemize} 

The modified Friedmann equation can be rewritten as follows.
\bea  \label{eq:H2a}
H^2 &=& \frac{\xi}{6}\left[ \frac{\dot{\phi}^2}{2} + V(\phi)+ \frac{ (\nabla \phi )^2}{2} +\rho _v \right]
- \frac{\xi ^2 \mu ^2 }{16} \Big[ \Lambda _\omega ^2 \nn \\ & -&  2 K \Lambda _\omega \left(\frac{\dot{\phi}^2}{2} + V(\phi)+ \frac{ (\nabla \phi )^2}{2} +\rho _v \right)^{\frac{2}{3(1+\omega)}} + 
 K^2  \left( \frac{\dot{\phi}^2}{2} + V(\phi)+ \frac{ (\nabla \phi )^2}{2} +\rho _v \right)^{\frac{4}{3(1+\omega)}}\Big], \hspace*{5mm}
\eea
where  $\rho_v=\Lambda \omega/8 \pi G$ represents gravitational energy density. 
The total energy density is composed of contributions from the inflationary potential and the cosmological ones, i.e. $\rho = \rho_{\phi} + \rho_{v}$.

In rapidly expanding Universe, the change in the inflationary field is likely homogeneous and very slow \cite{Liddle:2003, Linde:2002}. This situation can be modelled by viscous medium in a sphere:
\begin{itemize}
\item for homogeneous and isotropic FLRW Universe
\bea
(\nabla \phi)^2 &\ll & V(\phi),  \label{eq:inql1}
\eea
\item when the scalar field changes very slowly, one can neglect the acceleration so that
\bea
\ddot{\phi} & \ll & 3\, H\, \dot{\phi}, \label{eq:inql2} 
\eea
\item and under the assumption that the expansion of the Universe is so conditioned that the kinetic energy should be much less than the potential energy,
\bea
\dot{\phi}^2 & \ll &  V(\phi). \label{eq:inql3} 
\eea
\end{itemize}
In all these cases, it was assumed that both matter and radiation energy densities $\rho _m$ and $\rho_r$, respectively, can be neglected \cite{Liddle:2003, Linde:2002}. When assuming that the cosmic geometry is filled up with cold dark energy, e.g. $\omega=-1/3$ and $-2/3$, the modified Friedmann equation, Eq. (\ref{eq:H2a}), can straightforwardly be obtained.

\subsection{HLG with detailed balance conditions}
\label{sec:HLGdb}

This was proposed as a technical treatment reducing the number of the coupling constants, but by the end of the day HLG with detailed balance conditions should be broken in order to enable the theory being compatible with the available observations \cite{Charm,Horava2011a}. The detailed balance conditions lead to a fourth order dispersion relation for the scalar graviton, and thus spoiling the power-counting renormalisable feature of HLG \cite{1112.3385}. 

Accordingly, the modified Friedmann equations with the continuity equation is given as \cite{Tawfik:2017ngn}, 
\bea
H^2 &=& \frac{2}{(3\lambda - 1)} \Big[ \frac{\Lambda _\omega}{2} +  \frac{8 \pi G_N}{3} \left( V(\phi) + \frac{\Lambda _\omega}{8\, \pi \, G} \right)   - \kappa  \left( V(\phi) + \frac{\Lambda _\omega}{8\, \pi \, G} \right)^{\frac{2}{3(1+\omega)}}   \nn \\
&+& \frac{k^2}{2 \Lambda _\omega} \left( V(\phi) +  \frac{\Lambda _\omega}{8\, \pi \, G} \right)^{\frac{4}{3(1+\omega)}} 
\Big]. 
\eea

\subsection{HLG without the projectability conditions}
\label{sec:wtproj}

The projectability conditions are taken with the balance conditions and assume that the lapse function in the Arnowitt-Deser-Misner (ADM) decompositions is time-dependent. It was a big challenge to attempt to restore Lorentz symmetry at low energies, i.e. no mechanism ensuring that all states of matter and gravity have the same speed. This can be illustrated when recalling spin-0 graviton, which causes difference of its speed from that of spin-2 graviton. The latter is apparently not depending on any symmetry. If the detailed balance conditions are softly broken, it is assumed that the symmetry of HLG should be enlarged to include local U($1$) symmetry \cite{Wang2011sds,Wang2012ss}. Such a proposal \cite{Wang2012ss,1007.1006,1004.2552} eliminated the spin-0 graviton and the coupling constant $\Lambda_g$ in flat FLRW space-time \cite{Wang2012ss,1007.1006,1004.2552}. At vanishing cosmological constant, 
\bea
d s^2 &=& a^2(\eta) \left[ -d\eta^2 + \gamma_{ij}  dx^i dx^j\right],
\eea
where $\eta=\int dt/a$ is the conformal time and $\gamma_{ij} = \delta_{ij}/(1+ \kappa r^2/4)$. 

The modified Friedmann equation can be given as
\bea
{^{\prime}H^2} = \frac{8\, \pi \tilde{G}\, a^2}{3}\, \left(\frac{1}{2} \hat{\phi}^{\prime\, 2} + \tilde{V}(\hat{\phi})\right), 
\eea
where ${^{\prime}H}=H \, a$, ${^{\prime}\phi}=\dot{\phi}\, a$ , $\tilde{G}=2\, f\, G/(3\, \lambda -1)$, the background scalar field $\hat{\phi}=\hat{\phi}(\eta)$ and inflation potential $\tilde{V}(\hat{\phi})=V(\hat{\phi})/f(\lambda)$. Then,
\bea
H^2 = \frac{16\, \pi\, f\, G}{3 (3 \lambda -1) }\, \left(V(\phi)+\frac{\Lambda_{\omega}}{8\, \pi\, G}\right). \label{eq:woprojct}
\eea
It is apparent that Eq. (\ref{eq:woprojct}) does not depend on EoS, as the brackets in right-hand side include only energy densities related to the scalar field and to the cosmological constant. Thus, at $\lambda=1$ and with one-to-one mapping, i.e. $f=1$, Eq. (\ref{eq:woprojct})  can be reduced to standard Friedmann equation, and the standard GRG is restored, as well.

\section{Results}
\label{sec:Fried}

Based on minimal supersymmetric extensions to the standard model for cosmology, an inflationary potential has been proposed \cite{allahverdi-2006}
\bea
V(\phi) = \left(\frac{m^2}{2}\right)\,\phi^2   - \left(\frac{\sqrt{2\,\lambda\,(n-1)}\,m}{n}\right)\, \phi^n  + \left(\frac{\lambda}{4}\right)\,\phi^{2(n-1)}.
\eea
While $n>2$ is an integer, this model has two free parameters, $m$ and $\lambda$. For instance, at $n=3$, 
 \bea
V(\phi) = \left(\frac{m^2}{2}\right)\,\phi^2   - \left(\frac{2\sqrt{\lambda}\,m}{3}\right)\, \phi^3  + \left(\frac{\lambda}{4}\right)\,\phi^{4}. \label{eq:mssm}
\eea
The main reason why this scalar field is chosen is apparently merely motivated by its ability to cope with recent cosmological observations, BICEP2/Keck Array-Planck. This shall be illustrated in Fig. \ref{rVsns}.

\begin{figure}[htb!]
\centering{
\includegraphics[width=5.cm,angle=-90]{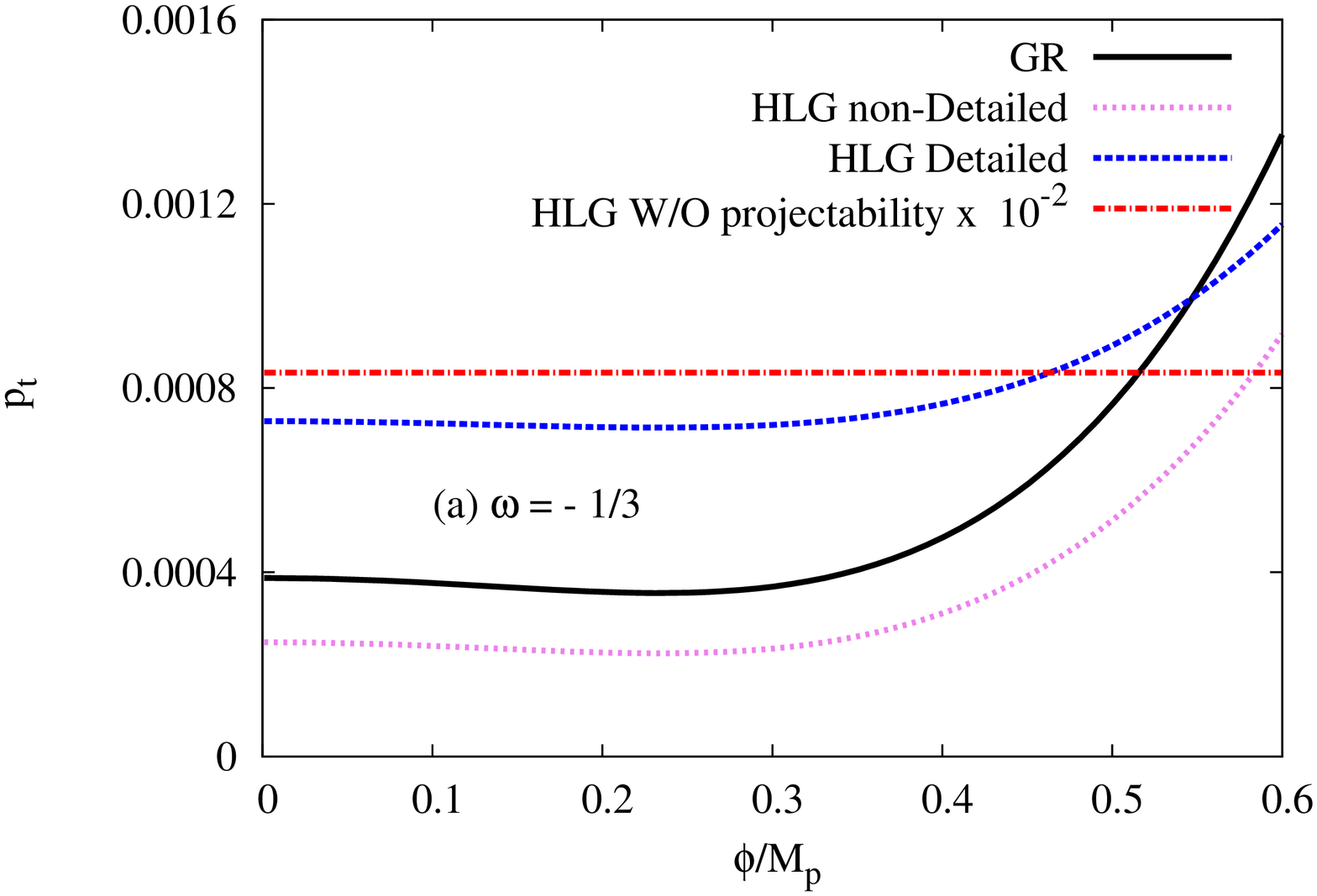}
\includegraphics[width=5.cm,angle=-90]{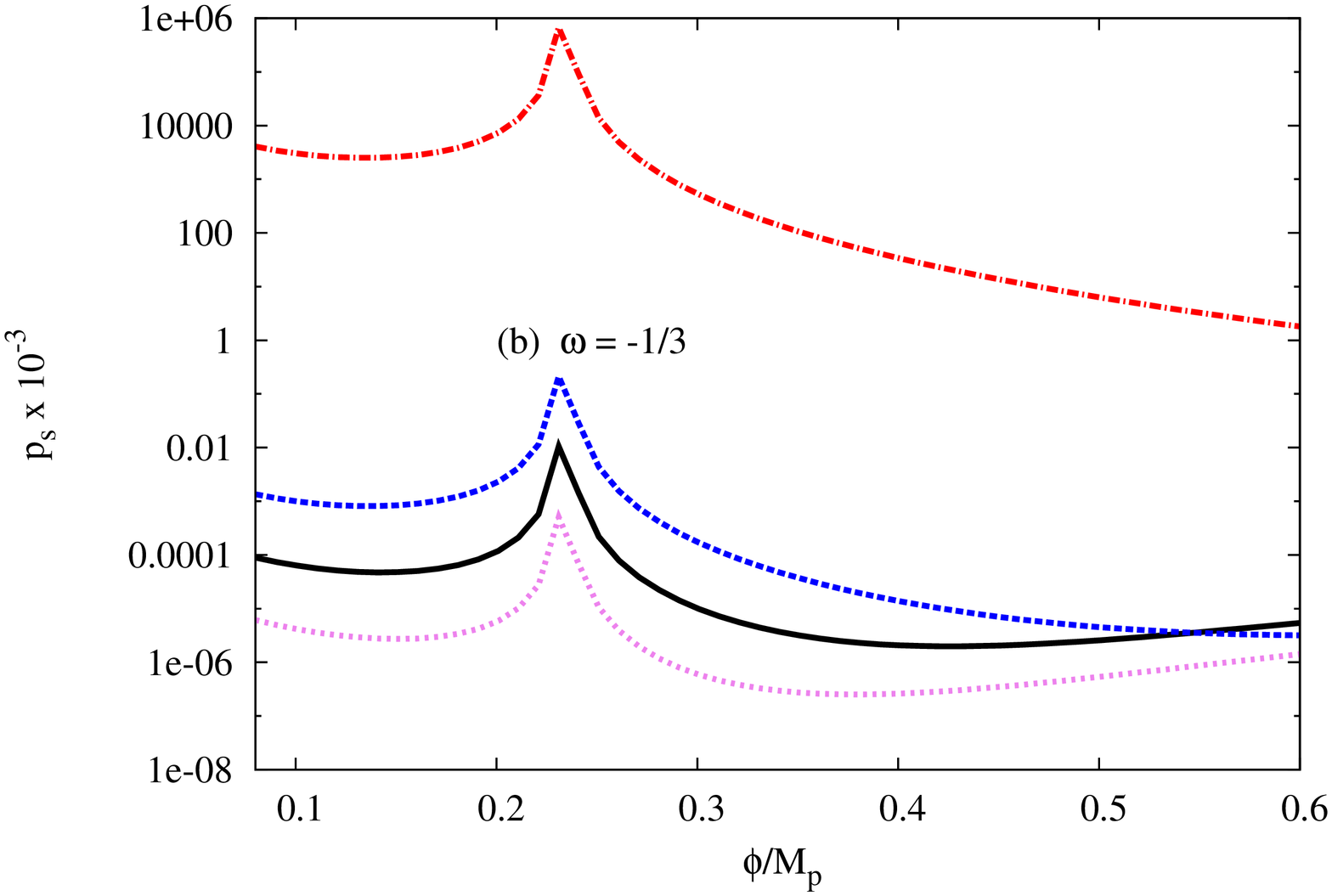} \\
\includegraphics[width=5.cm,angle=-90]{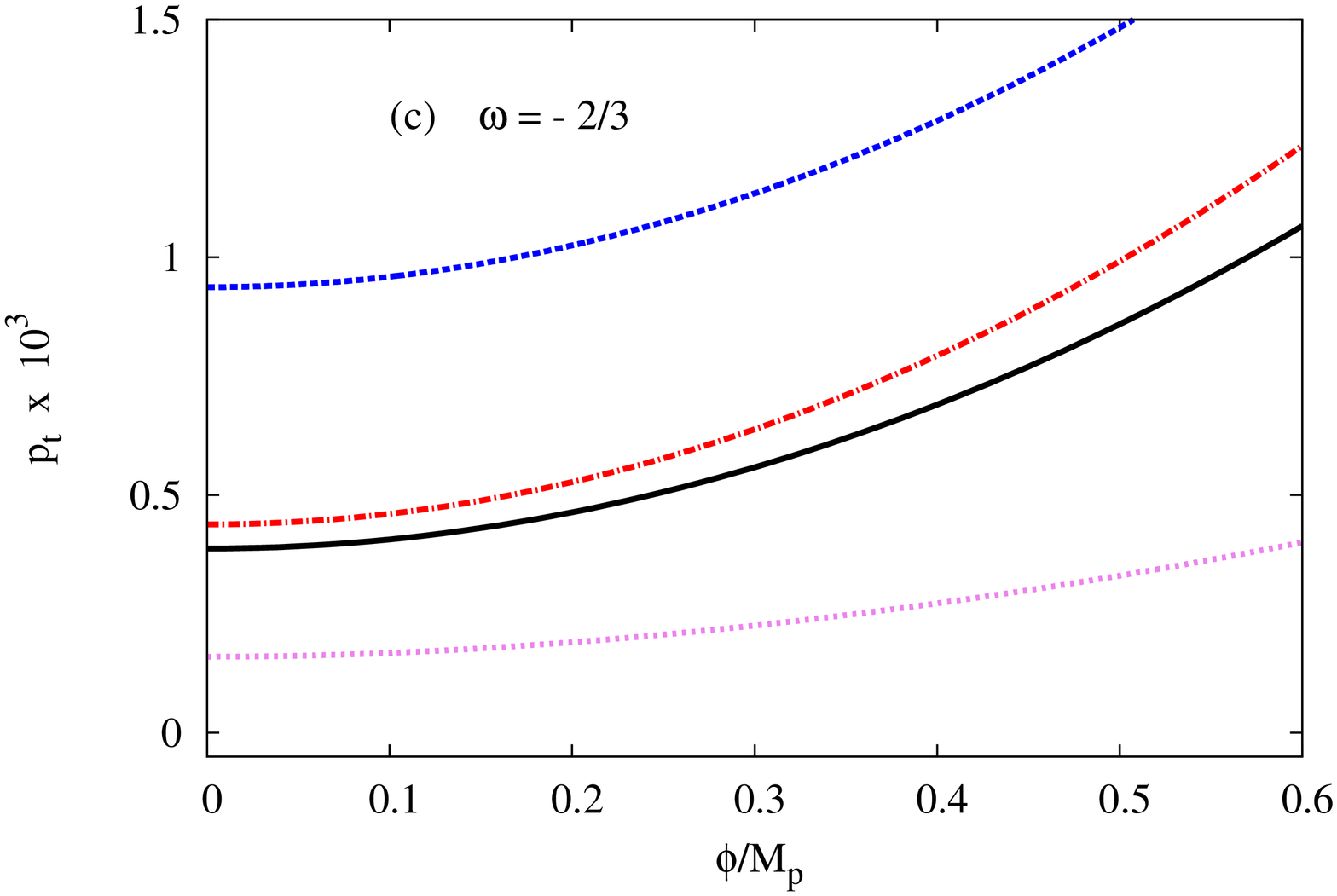}
\includegraphics[width=5.cm,angle=-90]{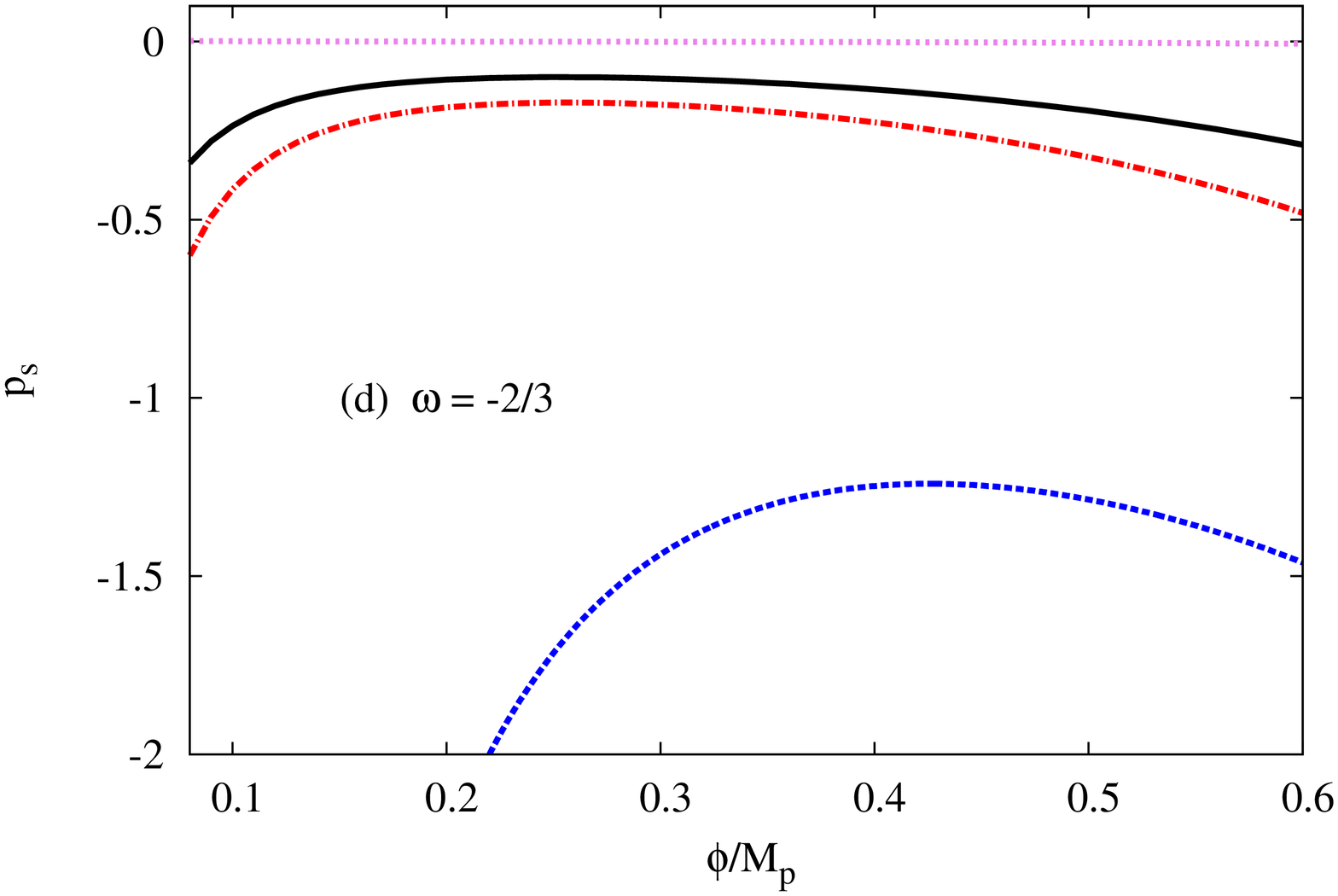}
\caption{Top-left panel: tonsorial density fluctuations ($P_t$)  from GRG (solid curve), HLG with detailed (dashed curve), HLG with non-detailed (dotted curve) and HLG without the projectability conditions (dash-dotted curve) are given as functions of inflation field ($\phi$) in units of the Planck mass $M_{p}$ for the proposed inflation potential at EoS $\omega=-1/3$. Top-right panel shows the same as in the top-left panel but for spectral density fluctuations ($P_s$).  Bottom-panel: the same as the top-panels but at the second EoS $\omega = -2/3$.
\label{tensorialfluc}
}}
\end{figure}

\begin{figure}[htb!]
\centering{
\includegraphics[width=5.cm,angle=-90]{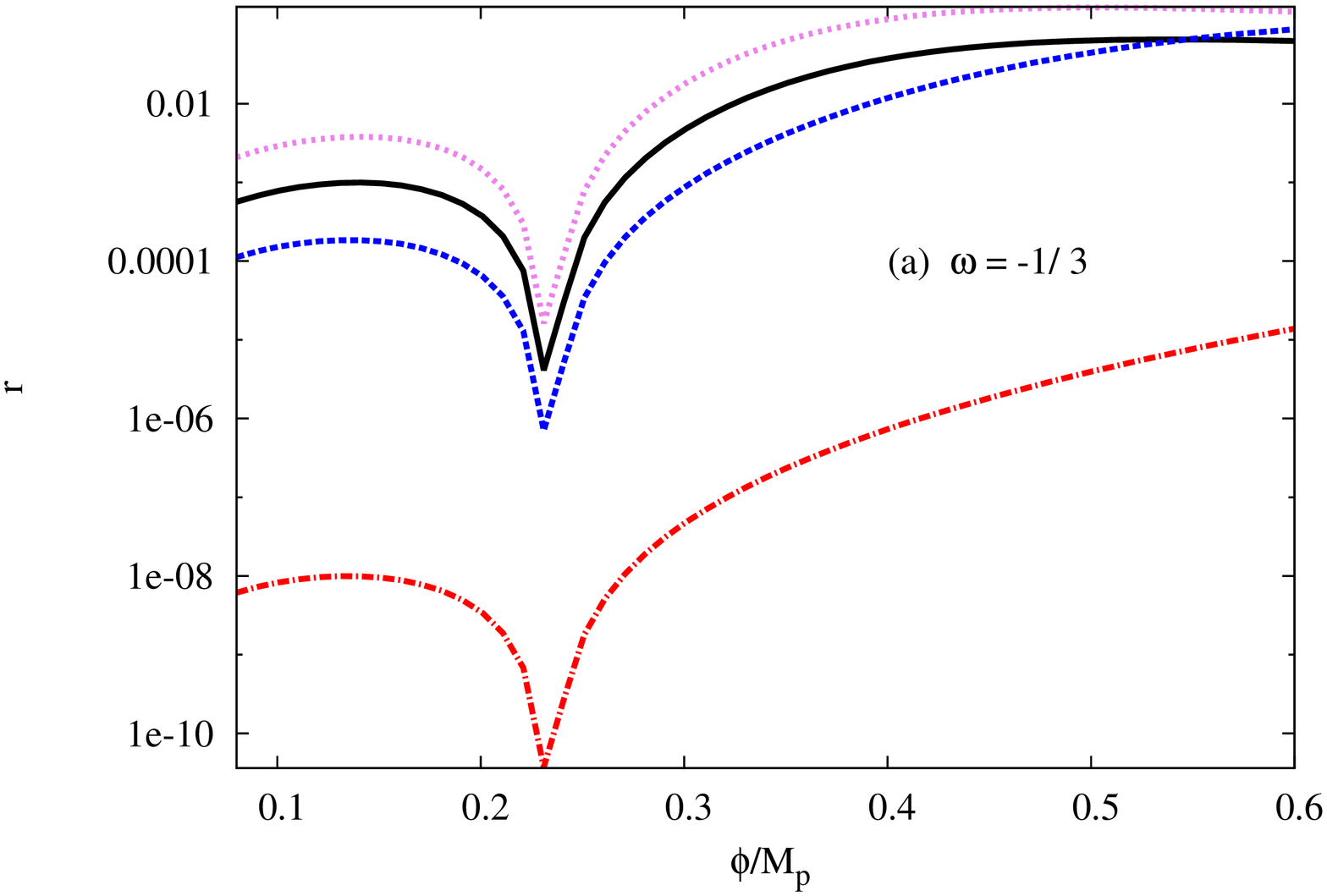}
\includegraphics[width=5.cm,angle=-90]{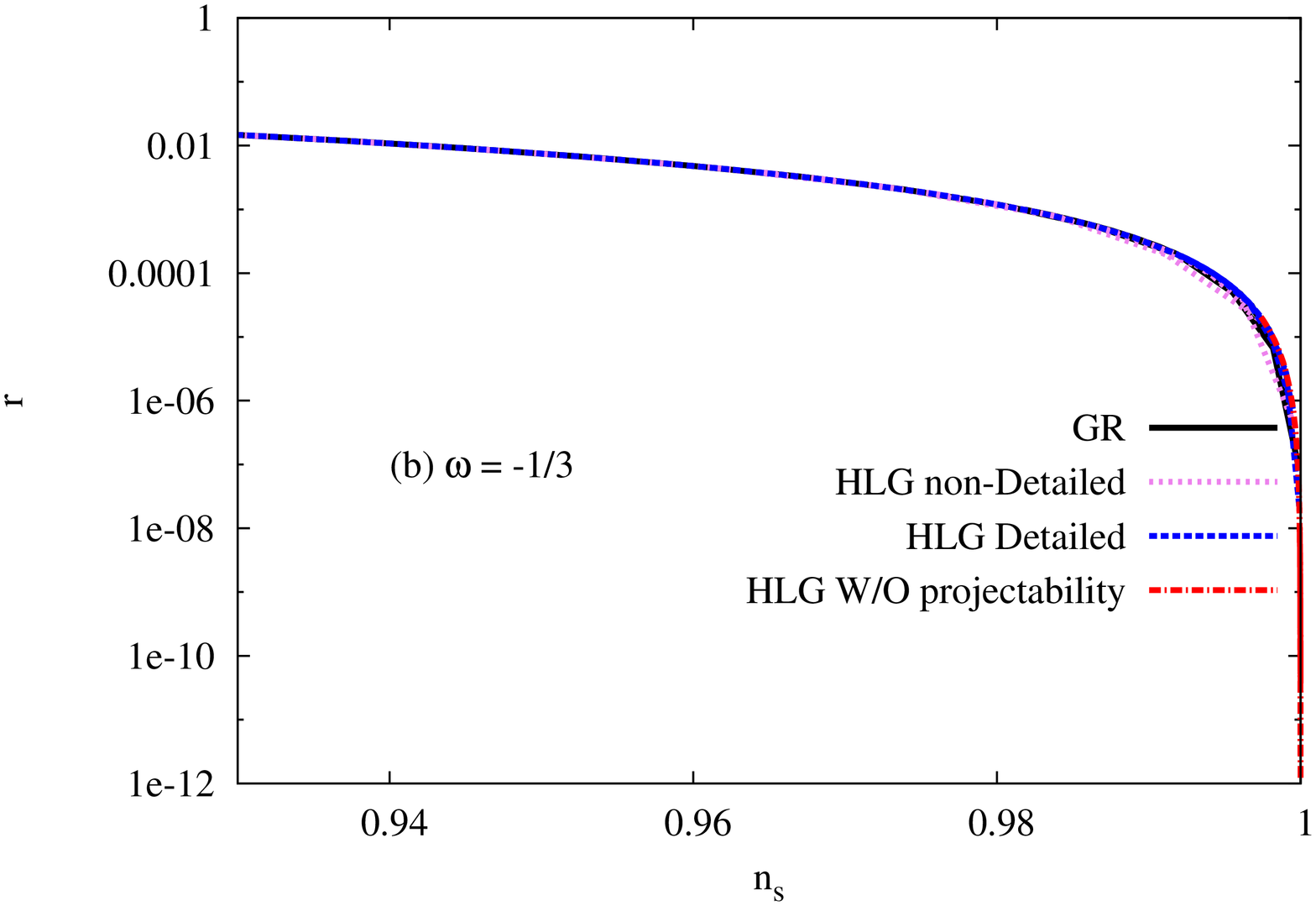}\\
\includegraphics[width=5.cm,angle=-90]{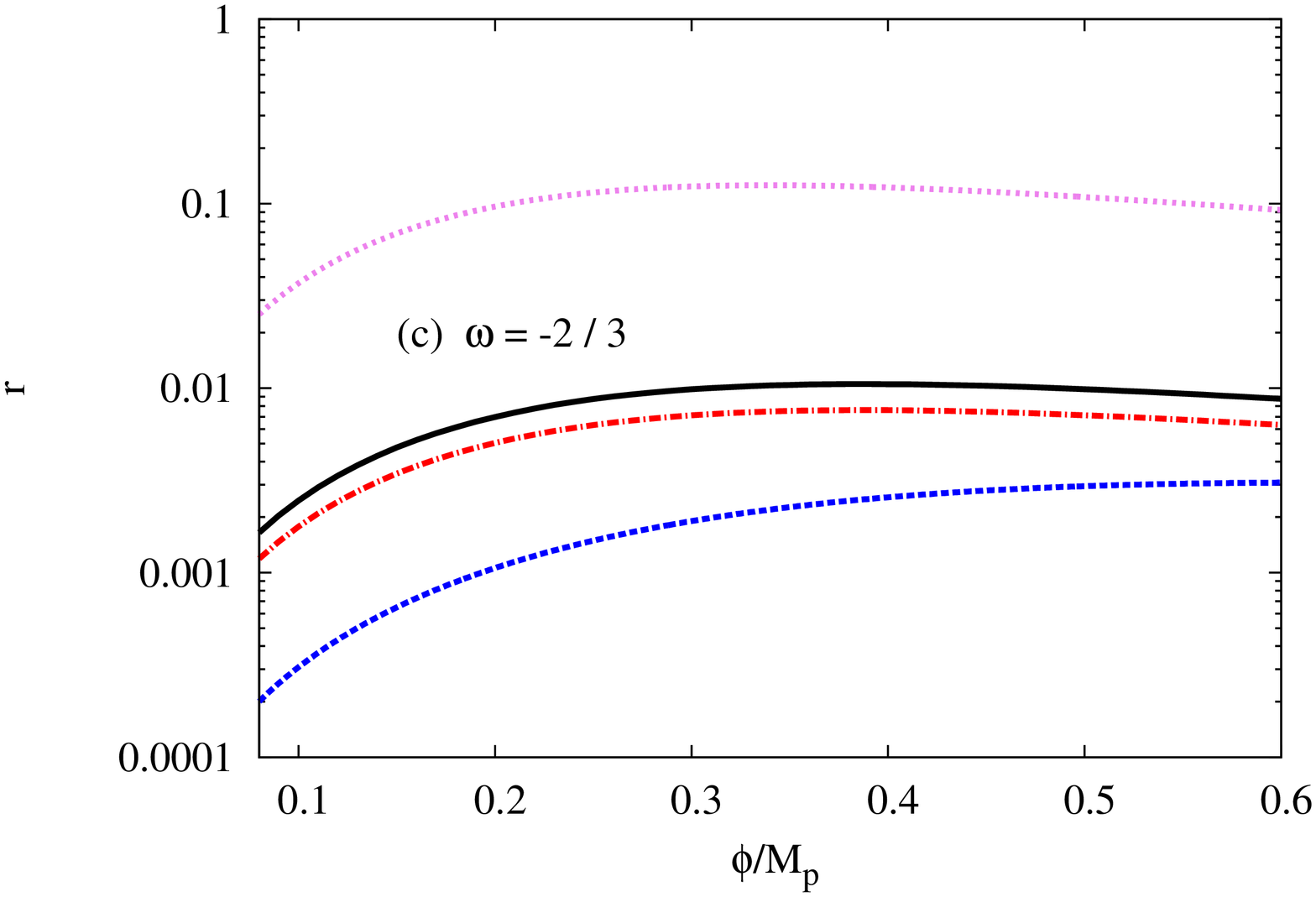}
\includegraphics[width=5.cm,angle=-90]{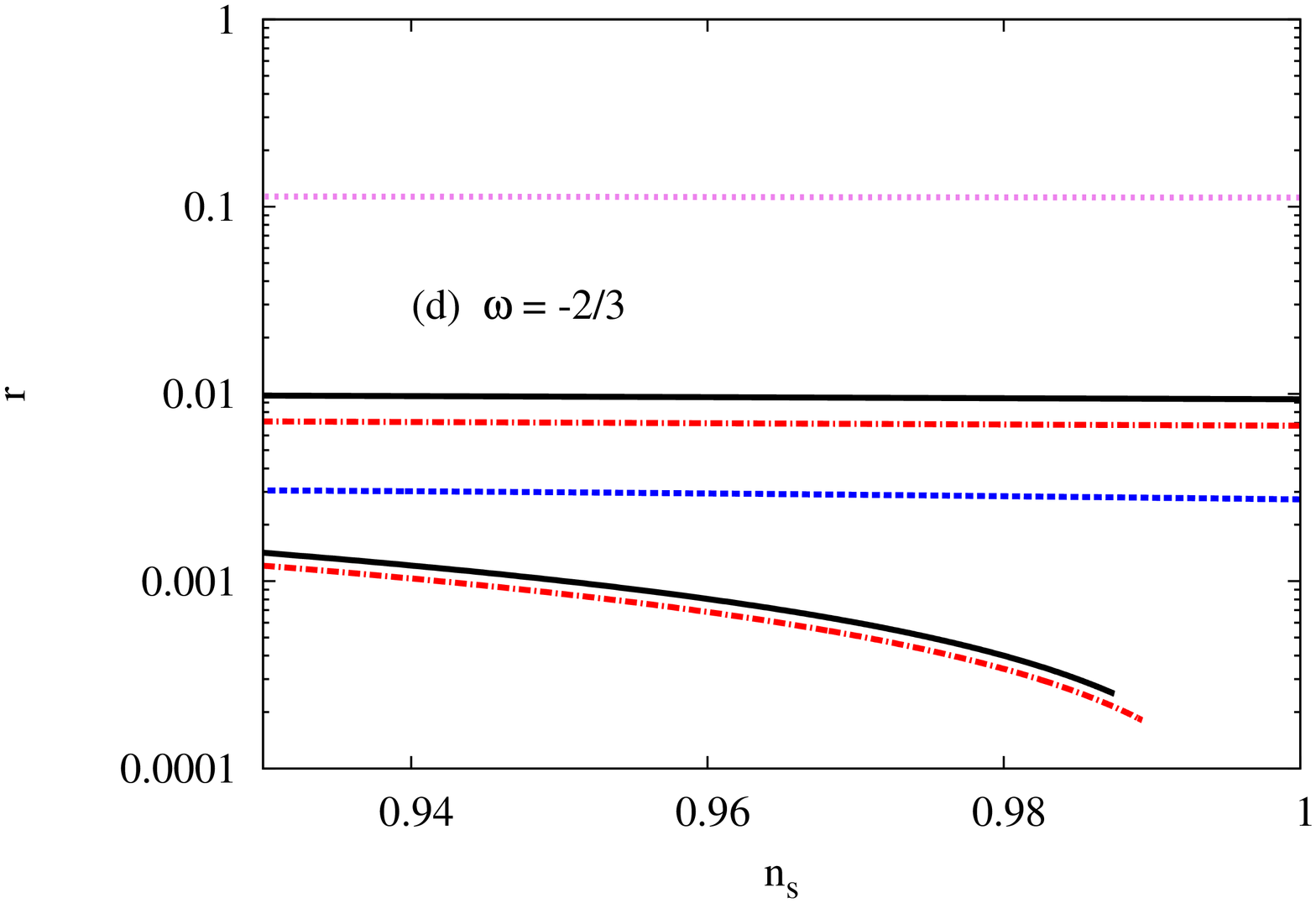}
\caption{The same as in Fig. \ref{tensorialfluc} but for tonsorial-to-spectral density fluctuations ($r=P_t/P_s$). Top-left panel: tonsorial-to-spectral density fluctuations ($r$) as a function of the inflation field ($\phi$) in units of the Planck mass ($M_{p}$), at EoS $\omega=-1/3$.  Top-right panel shows the same as in the top-left panel but here for tonsorial-to-spectral density fluctuations ($r$) as a function of the spectral index ($n_s$).  Bottom-panel gives the same as the top-panel but at EoS $\omega=-2/3$.
\label{tensorialspectralfluc}
}}
\end{figure}

\subsection{Tensorial and scalar density fluctuations} 
\label{FluctuationHLG}

It was pointed out that, while the scaler field ($\phi$) generates inflation \cite{Linde:1982,Liddle:1993,Liddle:1995}, the detailed balance conditions are conjectured not being necessary for the inflation \cite{0904.2190}. Nevertheless, the inclusion of a scalar field remains essential even in HLG.  This is the main difference between the present work and Ref. \cite{Mukohyama2010}. A comprehensive comparison shall be outlined in a future work. The main potential slow-roll parameters can be given as follows.
\bea
\epsilon _V  &\equiv & \frac{M_{p}^2}{16 \, \pi} \left(\frac{\partial _{\phi} V(\phi)}{ V(\phi)}\right)^2,
\label{paramters1} \\
\eta _V & \equiv & \frac{M_{p}^2}{8 \pi} \left(\frac{\partial_{\phi}^2 V(\phi)}{V(\phi)}\right),
\label{paramters2}
\eea
where $V(\phi)$ describes an inflation potential, such as Eq. (\ref{eq:mssm}). Moreover, the study of tensor-to-scalar fluctuations \cite{Linde:1982,Liddle:1993,Liddle:1995}
\bea
r &=& \frac{P_t}{P_s }=\left(\frac{\dot{\phi}}{H}\right)^2, \label{eq:r}
\eea
is more convenient as its dependence on various parameters turns to be eliminated. The first and second time evolutions of the inflation field, respectively, are given as 
\bea
\dot{\phi} &=& \frac{-1}{3H} \partial_{\phi} V(\phi), \\ 
\ddot{\phi} &=&  \frac{- \dot{\phi}}{3 H} \partial ^2_{\phi} V(\phi).
\eea
It is conjectured that the inflation with quantum fluctuations represents a mechanism to explaining the structure of the Universe. Thus various perturbations can be studied \cite{Wagenaar2016,Liddle:199,Liddle1994sa, Steinhardt1984ss,Liddle1992,Copeland1993s, Gong2009f}. First, scalar density perturbations are given by scalar functions of space-time coordinates. These are found closely related to growing density perturbations and enable to constrain the energy scale of inflation. Second, vector (tensor) density perturbations are related to vorticity perturbations. These could have imprints on the gravitational waves.  The tensorial and scalar density fluctuations, respectively, can be expressed as \cite{Liddle:1993,Liddle:1995,Gong2009f}
\begin{eqnarray}
P_t &=& \left(\frac{H}{2\pi}\right)^{2} \left[1-\frac{H}{\Lambda}\,\sin\left(\frac{2\Lambda}{H}\right)\right], \label{eq:Pt}\\
P_s &=& \left(\frac{H}{\dot{\phi}}\right)^2\left(\frac{H}{2\pi}\right)^{2} \left[1-\frac{H}{\Lambda}\,\sin\left(\frac{2\Lambda}{H}\right)\right]. \label{eq:Ps} \hspace*{10mm}
\label{ps}
\end{eqnarray}

Fig. \ref{tensorialfluc} illustrates the tonsorial ($P_t$) and scalar ($P_s$) density fluctuations as functions of $\phi$ normalized by the Planck mass $M_{p}$, Eq. (\ref{eq:Pt}) and (\ref{eq:Ps}), respectively, for the inflation potential, Eq. (\ref{eq:mssm}) at EoS characterized by $\omega=-1/3$ (top-left panel) and $-2/3$ (bottom-left panel). Our calculations are performed with GRG (solid curve), HLG with detailed (dashed curve), HLG with non-detailed (dotted curve) and HLG without the projectability (dash-dotted curve) conditions. We notice that the variation of density fluctuations with $\phi/M_{p}$ depends on: 
\begin{enumerate}
\item the proposed inflation potentials, 
\item the EoS characterizing the cosmic background as reported in Ref. \cite{Tawfik:2016dvd}, and 
\item the theories of gravity.
\end{enumerate}
Positive $P_t$ means that the second term in Eq. (\ref{eq:Pt}) is either positive but also it might be less than one or equivalently $\Lambda \lessapprox (\pi/2)\, H$. 

It is obvious that for GRG, and HLG with detailed and non-detailed balance conditions, $P_t$ remains constant, especially at low $\phi/M_{p}$.  At $\phi/M_{p}\gtrapprox 0.3$, $P_t$ increases with increasing  $\phi/M_{p}$. For HLG without the projectability conditions, although the resulting $P_t$ is relatively large, it is not depending on $\phi/M_{p}$. For a better comparison, $P_t$ from HLG without the projectability conditions is scaled by $10^{-2}$.

\begin{figure}[htb!]
\centering{
\includegraphics[width=5.cm,angle=-90]{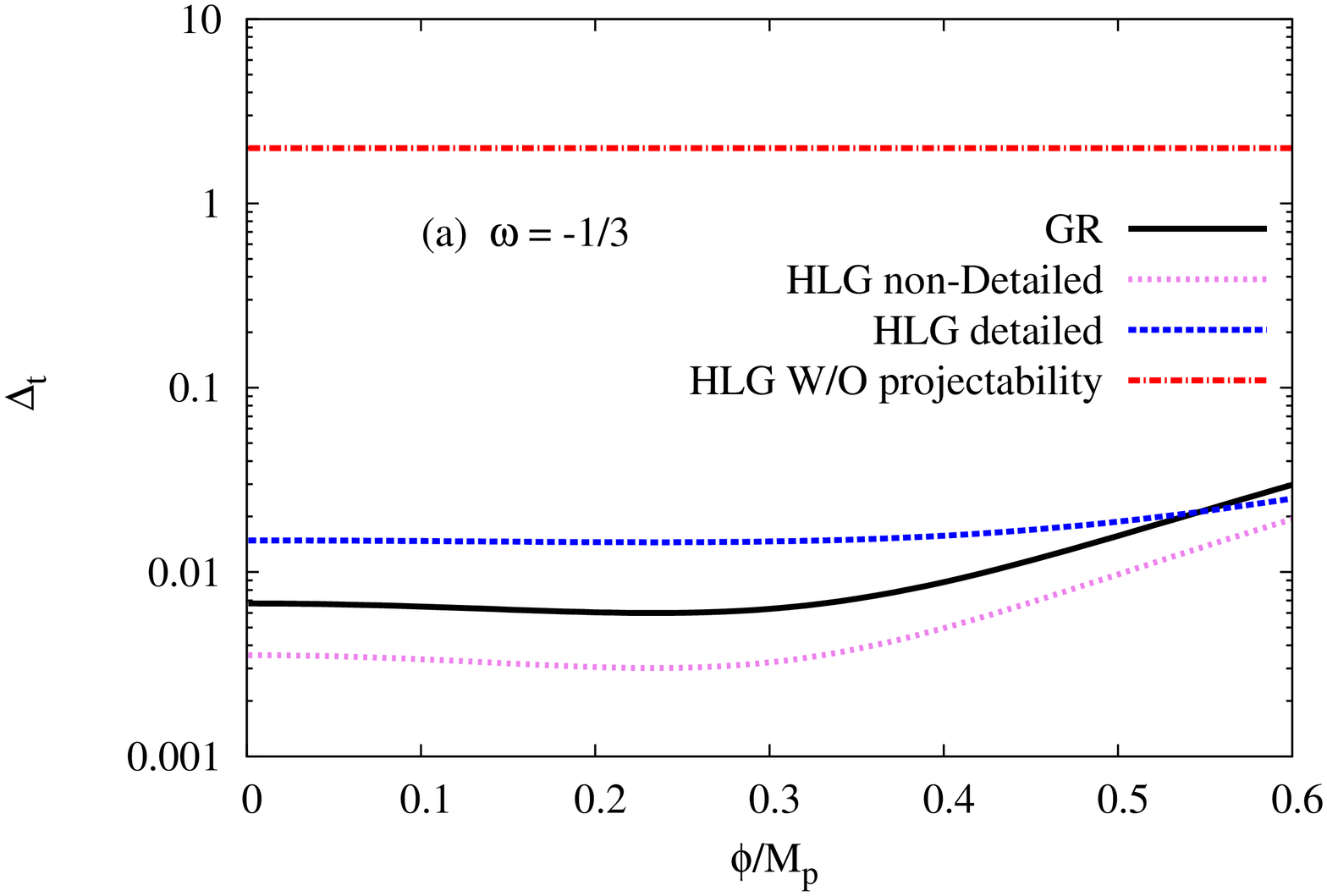}
\includegraphics[width=5.cm,angle=-90]{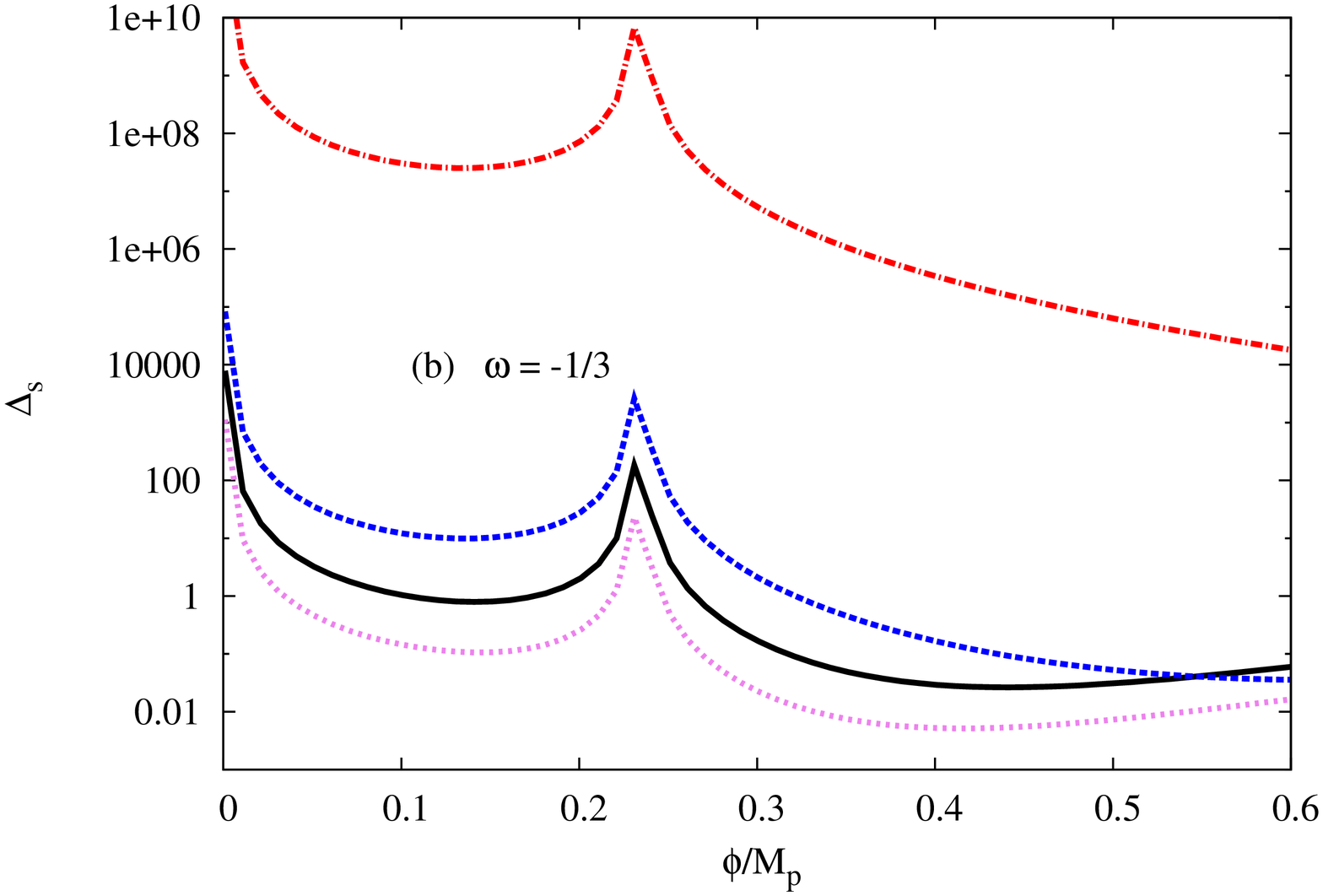}\\
\includegraphics[width=5.cm,angle=-90]{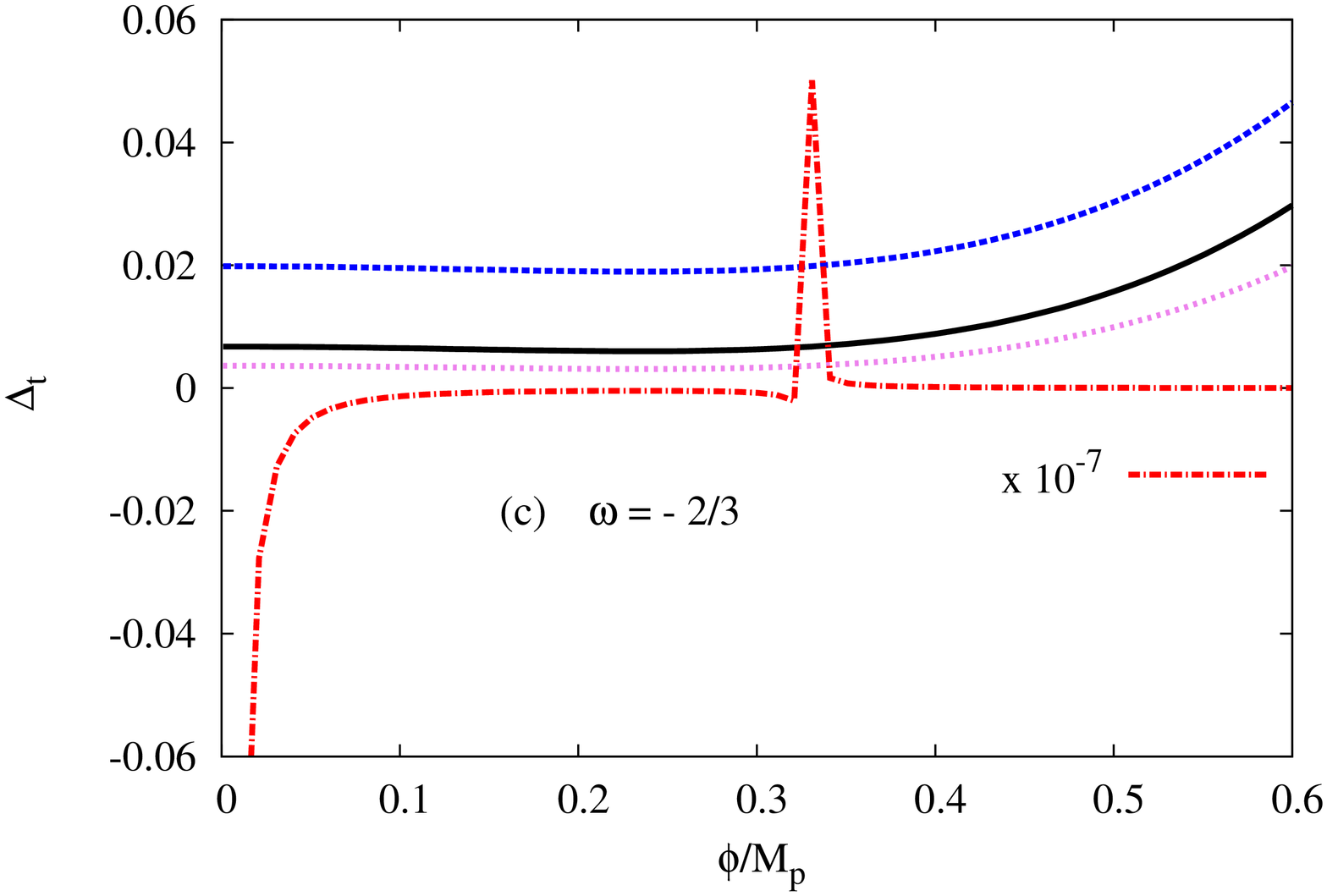}
\includegraphics[width=5.cm,angle=-90]{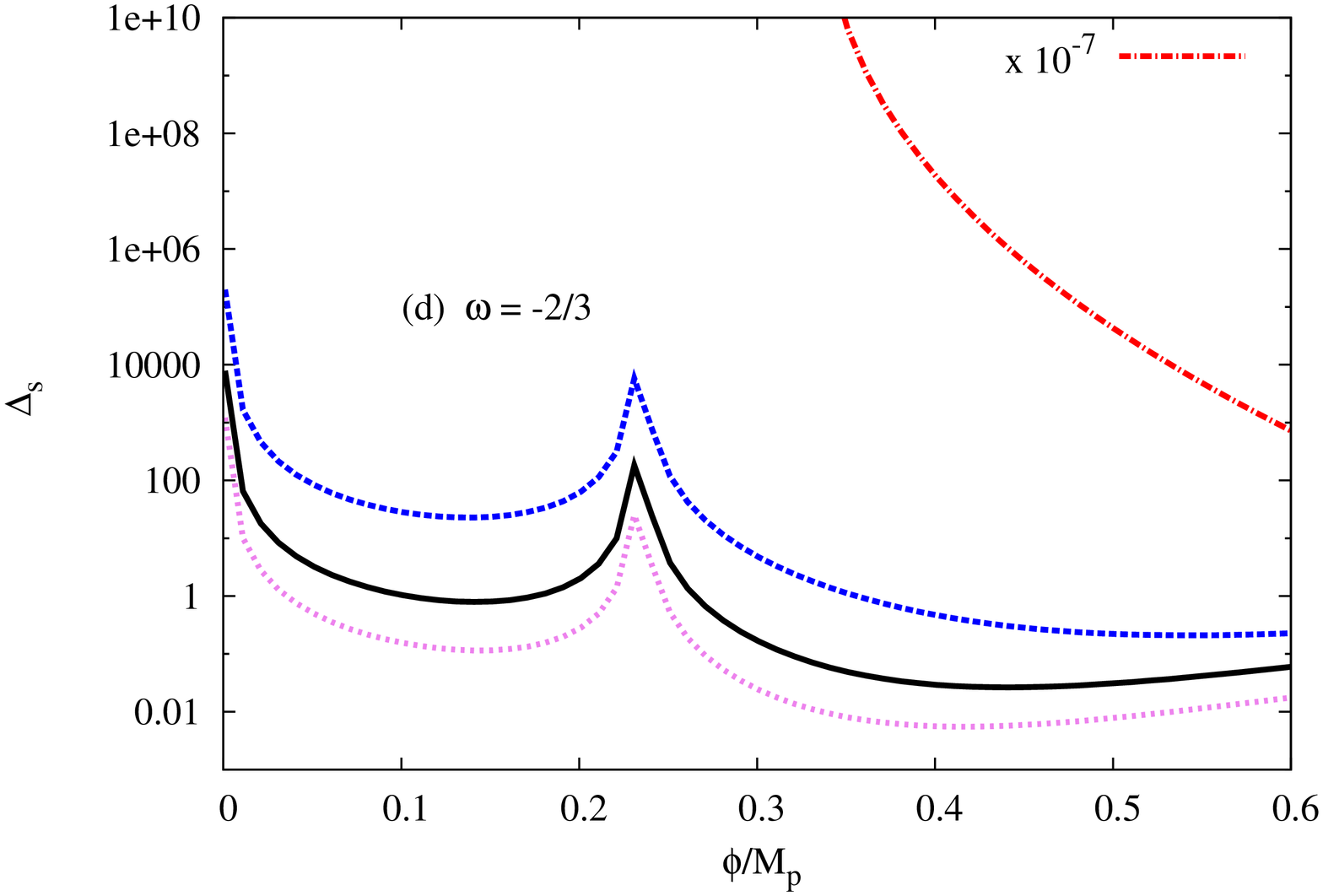}
\caption{Top-left panel: the primordial spectra of tensorial perturbations ($\Delta_t$)  from GRG (solid curve), HLG with detailed (dashed curve), HLG with non-detailed (dotted curve) and HLG without the projectability conditions (dash-dotted curve) are given as functions of the inflation field ($\phi$) in units of the Planck mass ($M_{p}$) for the proposed inflation potential at  $\omega=-1/3$. Top-right panel shows the same as in top-left panel but for primordial spectra of scalar perturbations ($\Delta_s$).  Bottom-panels draws the same as the top-panels but at $\omega=-2/3$.
\label{tensorialprimordial}
}}

Top-right panel of Fig. \ref{tensorialfluc} depicts the spectral density fluctuations ($P_s$), Eq. (\ref{eq:Ps}), versus $\phi/M_{p}$. Almost similar conclusions can be drawn. This allows us to conclude a dominant contribution from the cosmological constant. Characteristic peaks are observed at $\phi/M_{p} \simeq 2.3$. Again, the resulting $P_s$ from HLG without the projectability conditions is much higher than the ones from GRG, and HLG with detailed and non-detailed balance conditions.

Top-left panel of Fig. \ref{tensorialspectralfluc} illustrates $r=P_t/P_s$, Eq. (\ref{eq:r}), the ratio of tensorial-to-spectral density fluctuations, as functions of $\phi/M_{p}$. There are characteristic peaks at $\phi/M_{p} \simeq 2.3$. We notice that $r$ from HLG without the projectability conditions is much less than the ones from GRG, and HLG with detailed and non-detailed balance conditions. It is obvious that the last three types of gravity theories are much compatible with each others. 

Top-right panel of Fig. \ref{tensorialspectralfluc} shows $r$ as functions of the spectral index ($n_s$), Eq. (\ref{eq:ns}). Apparently, the results are not depending on the gravity theory.

\subsection{Tensorial and scalar perturbations of primordial spectra \label{sec:primrdlPertr}}

\end{figure}
\begin{figure}[htb!]
\centering{
\includegraphics[width=5.cm,angle=-90]{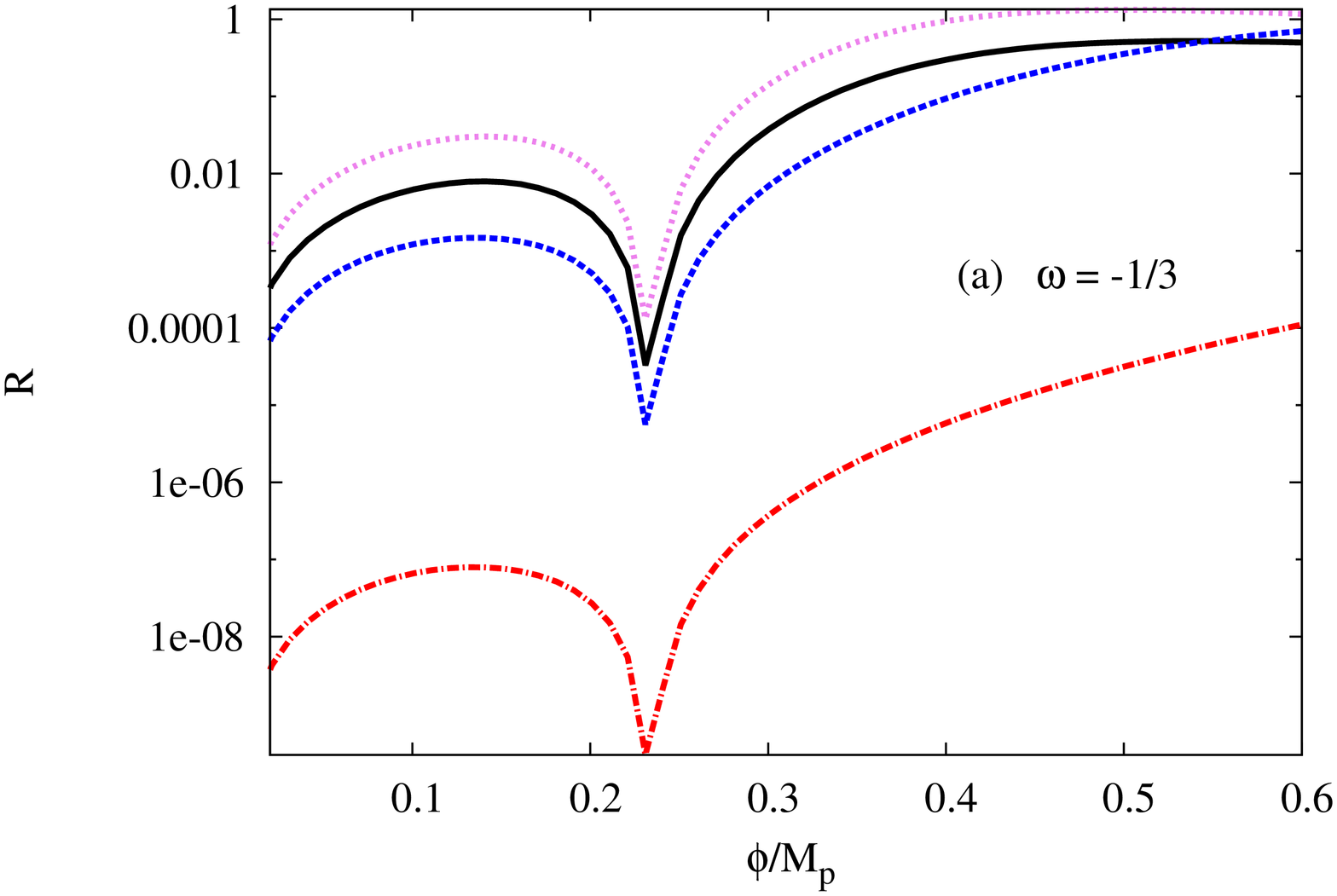}
\includegraphics[width=5.cm,angle=-90]{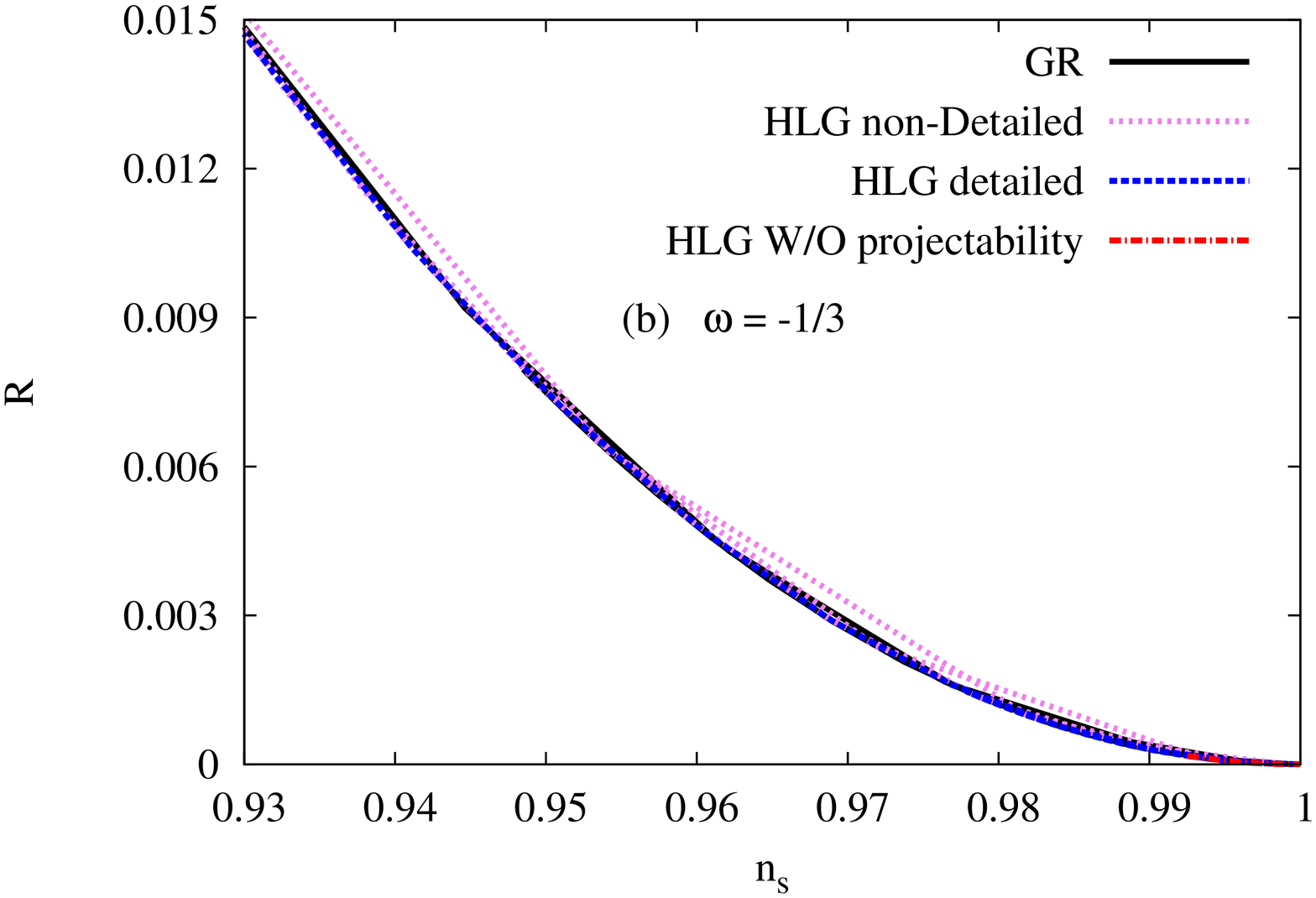}\\
\includegraphics[width=5.cm,angle=-90]{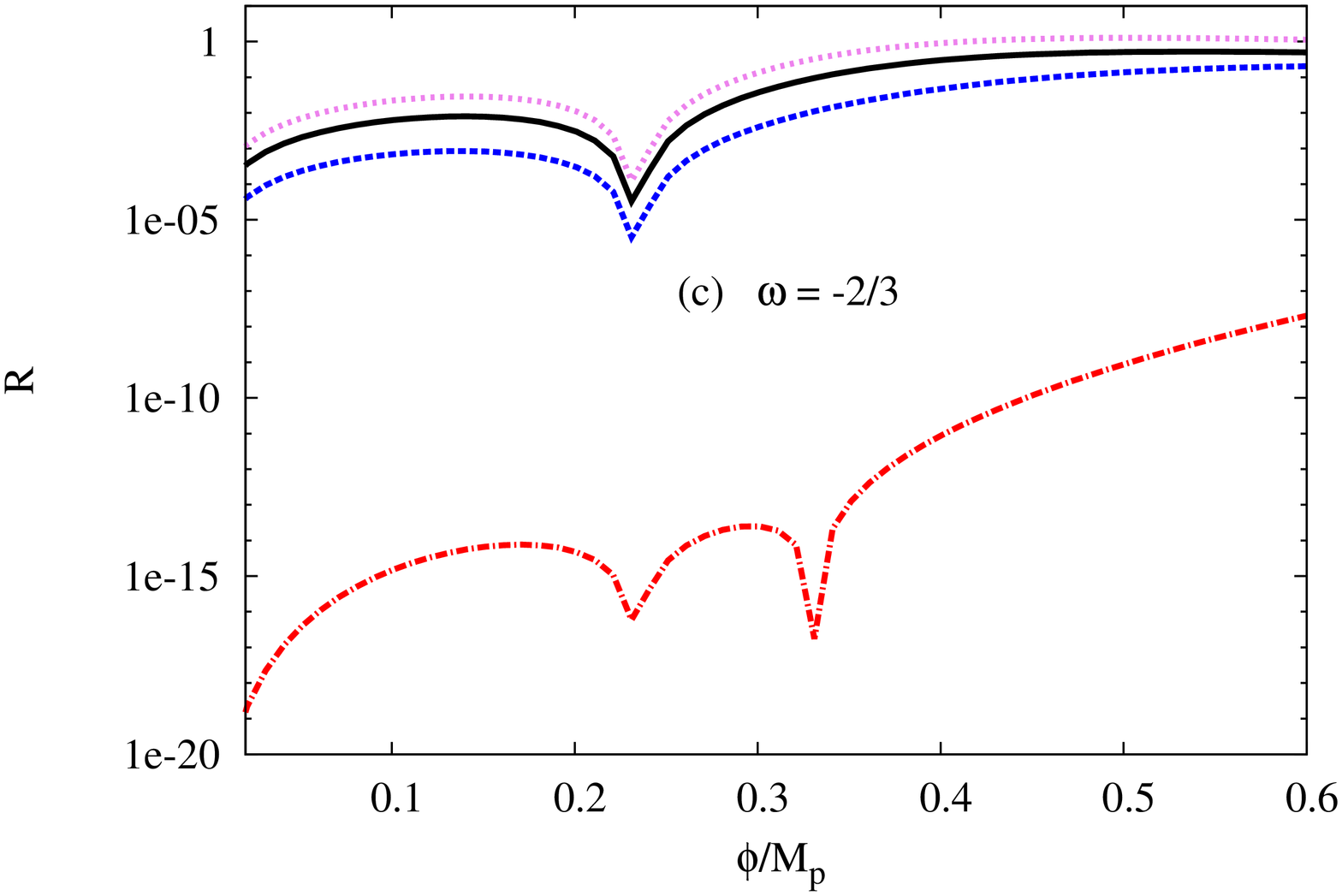}
\includegraphics[width=5.cm,angle=-90]{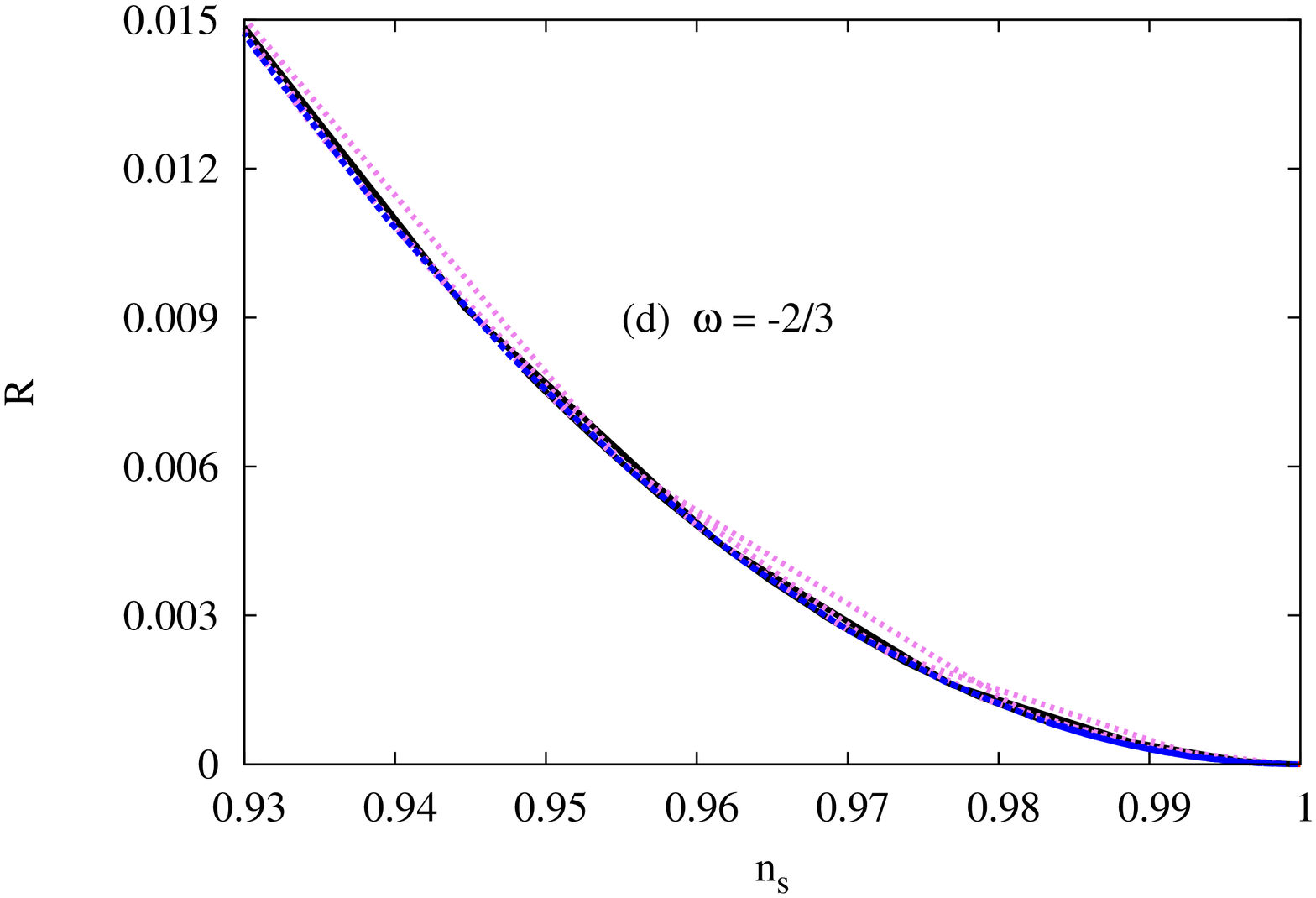}
\caption{The same as Fig. \ref{tensorialfluc} but for the ratio tonsorial-to-spectral perturbations primordial spectra  ($R=\Delta_t/\Delta_s$). Top-left panel: $R$ is given as a function of the inflation field ($\phi$) normalized to the Planck mass ($M_{p}$) at $\omega=-1/3$.  Top-right panel shows the same as in top-left panel but for $R$ as function of the spectral index ($n_s$).  Bottom-panel: the same as the top-panel but at $\omega=-2/3$.
\label{tensorialspectralprimordial}
}}
\end{figure}

Another alternative to compare our calculations with are the recent Planck observations. Concretely, the 
phenomenological parametrization of the primordial spectra of tensorial and scalar perturbations, which can be respectively given as
\begin{eqnarray}
\Delta_t (\phi) &=& \frac{2 H^2}{\pi^2 M_{p}^2} \sim \frac{2 V(\phi)}{3\pi^2 M_{p}^4}, \\
\Delta_s (\phi) &=& \left(\frac{H}{2\pi}\right)^2 \frac{H^2}{\dot{\phi}^2} = \frac{V(\phi)}{24\pi^2 M_{p}^4\; \epsilon_H}, 
\end{eqnarray}
where $\epsilon_H=2 M_{p}^2 (H_1/H_0)^2$ denotes the first Hubble slow-roll parameter, where the shape of the function $H(\phi)$ is conjectured to be well observed, and $H_0$ and $H_1$ are Taylor coefficients.

One can relate the primordial spectra to the time of the horizon crossing $\zeta = a\,H$, such as 
\bea
\Delta_t (\zeta)&=& A_t \left(\frac{\zeta}{\zeta_*}\right)^{n_t + \frac{1}{2} \frac{d n_t}{d \ln{\zeta}} \ln{\left[\frac{\zeta}{\zeta_*}\right]}+ \cdots}, \label{eq:Dltat}
\\ 
\Delta_s (\zeta)&=& A_s \left(\frac{\zeta}{\zeta_*}\right)^{n_s - 1 + \frac{1}{2} \frac{d n_s}{d \ln{\zeta}} \ln{\left[\frac{\zeta}{\zeta_*}\right]}+ \cdots}. \label{eq:Dltas}
\eea
where $\zeta_*$ is the pivot scale and $A_s$, $A_t$, $n_s$, $n_t$, $d n_s/ d \ln{\zeta}$ and $d n_t/d \ln{\zeta}$  are scalar amplitudes, tensor amplitude, scalar spectral index, tensor  spectral index, the running of the scalar (tensor) spectral index and the running of the tensor spectral index, respectively. The scalar and tensor spectra index can be given as 
\bea
n_s &\equiv & 1 - \sqrt{\frac{R}{3}},\\ \label{eq:ns}
n_t &=& -\frac{R}{8}\left(2-\frac{R}{8}-n_s\right),
\eea
with 
\bea
R &=& \frac{\Delta_t}{\Delta_s}, \label{eq:R}
\eea 
is to be fixed to $-8n_t$ and thus $d n_t/d\ln \zeta=R(R/8+n_s-1)/8$.

Top-panel of Fig. \ref{tensorialprimordial} illustrates the primordial spectra of tensorial, $\Delta_t$, Eq. (\ref{eq:Dltat}) (left-hand panel) and scalar perturbations of primordial spectra, $\Delta_s$, Eq. (\ref{eq:Dltas}), (right-hand panel), as functions of $\phi/M_{p}$, at $\omega=-1/3$. We observe that, GRG (solid curve), and HLG with detailed (dashed curves) and non-detailed balance conditions (dotted curves) are compatible with each others. At low $\phi/M_{p}$, $\Delta_t$ remain constant, while $\phi/M_{p}$ increases. At $\phi/M_{p}\simeq 0.3$, $\Delta_t$ very slowly increases with increasing $\phi/M_{p}$. HLG without the projectability conditions (dot-dashed curves) results in relatively large $\Delta_t$, which in turn is no depending on $\phi/M_{p}$. 

In bottom-panel of Fig. \ref{tensorialprimordial}, the same as in top-panel but at $\omega=-2/3$. For $\Delta_s$, there are characteristic peaks at $\phi/M_{p}\simeq 2.3$. Also, at vanishing $\phi/M_{p}$, there a rapid decrease in $\Delta_s$.

Top-right panel of Fig. \ref{tensorialspectralprimordial} presents the ratios of tonsorial-to-spectral perturbations density fluctuations of the primordial spectral ($R$), Eq. (\ref{eq:R}), as functions of $\phi/M_{p}$, at $\omega=-1/3$. Also here, the results from GRG (solid curve), and HLG with detailed (dashed curves) and non-detailed balance conditions (dotted curves) are found compatible with each others. They all have two minimals; one at vanishing $\phi/M_{p}$ and one at $\phi/M_{p}\simeq 0.23$. Again, the results from HLG without the projectability conditions (dot-dashed curves)  are much smaller than the ones from the other types of gravity. While, the bottom-left panel depicts $R$ vs. $n_s$. It is obvious that the results are not depending on the gravity. Bottom-right panel gives the same as in the top-panel but at $\omega=-2/3$.
\vspace*{5mm}

\subsection{BICEP2/Keck Array-Planck Observations \label{sec:planck} }

\begin{figure}
\begin{center}
\includegraphics[width=10.cm,angle=-0]{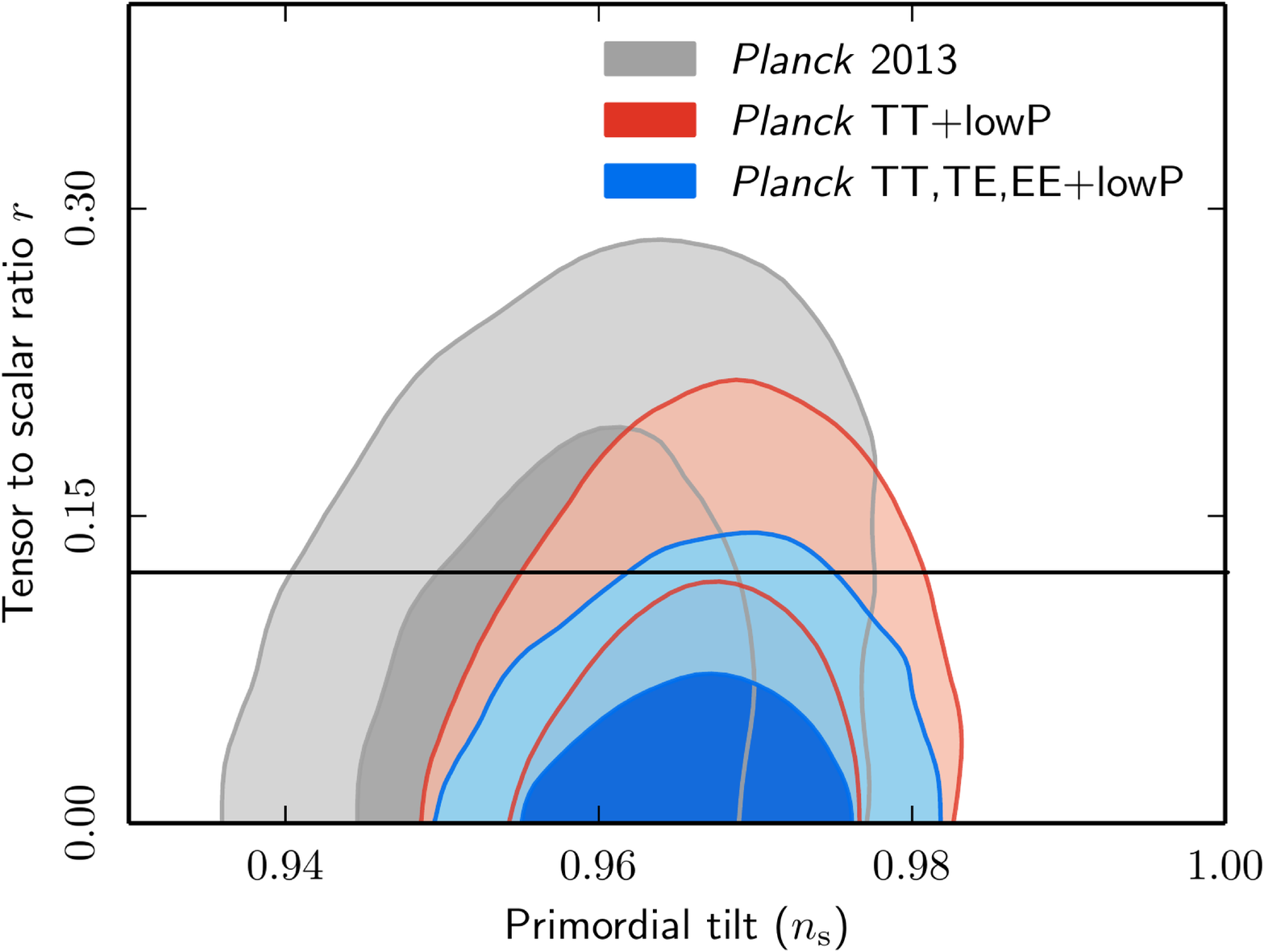} 
\caption{Results on $r$  vs. $n_s$ as plotted in top-right panel of Fig. \ref{tensorialspectralfluc} are confronted to the recent BICEP2/Keck Array-Planck observations (red and blue contours), which are also compared to previous observations (gray contours). The horizontal line represents our calculations at $\omega=-2/3$ in HLG with non-detailed balance conditions.}
\label{rVsns}
\end{center}
\end{figure}

In top-right panel of Fig. \ref{tensorialspectralfluc}, the tonsorial-to-spectral density fluctuations ratios ($r$) are given as functions of the spectral index ($n_s$). Both quantities have been analysed by the BICEP2/Keck Array-Planck collaborations. In Fig. \ref{rVsns}, we confront our calculations to the recent BICEP2/Keck Array-Planck observations. 

In Fig. \ref{rVsns}, the dependence of $r$ on $n_s$, as deduced from our parametric estimation, is confronted to BICEP2/Keck Array-Planck marginalized joint contours at $68\%$ and $95\%$ confidence level \cite{planck2015a,planck2015b}. The figure also illustrates a significant improvement (red and blue contours) with respect to previous BICEP2/Keck Array-Planck data release (gray contours). 

Recent BICEP2/Keck Array-Planck observations set an upper bound to the tensor-to-scalar ratio, $r_{0.002}<0.11$  at $95\%$ confidence level, when taking into consideration the BICEP2/Keck Array-Planck high-$\ell$ polarization data. The upper limit according to {\it B}-mode polarization constraint, $r<0.12$ at $95\,\%$ at confidence level, which was obtained from a joint analysis of BICEP2/Keck Array-Planck data \cite{planck2015a} is well reproduced.  

In Fig. \ref{rVsns}, $r$ as obtained from our parametric calculations (solid line) is determined from HLG with non-detailed balance conditions from the scalar field, Eq. (\ref{eq:mssm}), at $\omega=-2/3$. Our calculations for $r$ in dependence in $n_s$ are depicted in top-right panel of Fig. \ref{tensorialspectralfluc}, where it is obvious that $10^{-9} \lessapprox r \lessapprox 10^{-3}$.

\section{Conclusions}
\label{sec:conc}
 
We have studied the dependence of the tensorial and spectral density fluctuations and the perturbations of the primordial spectra on the scaler field characterizing a minimal-supersymmetric inflation field in Horava-Lifshitz gravity. The proposed scalar field is conjectured to generate the inflation. It was found that the proposed inflation potential, the cold dark energy EoS characterizing cosmic background geometry, and various gravity theories determine how density fluctuations and primordial spectra vary with the scaler field and with the spectral index. The dependence of tensorial-to-spectral density fluctuations on the quantum perturbations in form of scalar spectra index has been calculated and compared to the recent PLANCK observations, i.e. $r_{0.002}<0.11$  at $95\%$ confidence level in BICEP2/Keck Array-Planck high-$\ell$ polarization data. The upper limit according to {\it B}-mode polarization constrains is that $r<0.12$ at $95\,\%$ at confidence level. This was obtained from a joint analysis of BICEP2/Keck Array-Planck data \cite{planck2015a}. Our parametric calculations at $\omega=-1/3$ agree well with these observations. 

We also found that, at $\omega=-1/3$, the tensorial-to-spectral density fluctuations and the perturbations of the primordial spectra are not depending on the underlying theory of gravity; GRG, HLG with detailed and non-detailed balance conditions, and HLG without the projectability conditions. The results of $r$ and $R$ in HLG with non-detailed balance conditions range between $0.0996$ and $1.0$, i.e. much higher than BICEP2/Keck Array-Planck observations, while in GRG, and HLG with detailed and non-detailed balance conditions, the resulting $r$ and $R$ are $10^{-9} \lessapprox r \lessapprox 10^{-3}$ and $0.0 \lessapprox R \lessapprox 0.015$, respectively. These are much smaller than the BICEP2/Keck Array-Planck observations.

\end{document}